\documentclass[a4,12pt]{article}

\usepackage{amssymb,cancel,amsmath,indentfirst}
\usepackage{hyperref}

\begin{document}

\title{\bf \Large Quantum Action Principle with Generalized Uncertainty Principle}
\vspace{3em}

\author{\vspace{1em}Jie Gu\\  \emph{\small State Key Laboratory of Automotive Safety and Energy,}\\ \emph{ \small  Tsinghua University, Beijing 100084, People's Republic of China}}

\date{$\;$}

\maketitle
\thispagestyle{empty}

\bigskip

\begin{abstract}
One of the common features in all promising candidates of quantum gravity is the existence of a minimal length scale, which naturally emerges with a generalized uncertainty principle, or equivalently a modified commutation relation. Schwinger's quantum action principle was modified to incorporate this modification, and was applied to the calculation of the kernel of a free particle, partly recovering the result previously studied using path integral.
\end{abstract}

\pagebreak

%========================================================================================================================================
%% main text
\section{Introduction}
\label{}
Quantizing gravity is one of the most important tasks of modern theoretical physics which, up to this date, has not yet found a consistent and satisfactory solution. Quantum gravity phenomenology has arisen which is devoted to describing gravity effects in quantum systems and finding responding possible experimental signatures.

One of the common features in all promising candidates of quantum gravity (such as the string theory\cite{Gross1988,Konishi1990}, loop quantum gravity \cite{Rovelli1995} and other approaches \cite{Scardigli1999}) is the existence of a minimal length scale (see \cite{Hossenfelder2013} for a review). Actually the idea of a minimal length appears in these theories as either a prediction or a necessary part. As we will see, this minimal length could be introduced by a generalized uncertainty principle, which has received increasing attention in the last decade.

Consider a system defined by coordinates $\{q_a\}$ and momenta $\{p_a\}$, $a=1,\ldots,n$. One of the most fundamental concepts in quantum mechanics is the Heisenberg uncertainty principle
\begin{equation}
\Delta q_a \Delta p_b \ge \frac12, \quad \left( \hbar=1 \right),
\end{equation}
which is consistent with the canonical commutation relation
\begin{equation}
\left[ q_a,p_b \right]=i\delta_{ab}.
\end{equation}

Even though the uncertainty principle imposes a constraint that we could not measure the position and momentum as precisely as we want simultaneously, it does not prevent us from measuring the position with increasing precision as long as the momentum gets larger. However, if we modify the commutation relation from the canonical one to
\begin{equation} \label{mcr}
\left[ q_a,p_b \right]=i \left( \delta_{ab}+ \beta p^2 \delta_{ab}+ 2 \beta p_a p_b \right),
\end{equation}
one obtains
\begin{equation}
\Delta {q_a}\Delta {p_a} \ge \frac{1}{2}\left( {1 + 3\beta \Delta p_a^2} \right),
\end{equation}
which is the generalized uncertainty principle (GUP), and thus
\begin{equation}
\Delta {q_a} \ge \frac{1}{2}\left( {\frac{1}{{\Delta {p_a}}} + 3\beta \Delta {p_a}} \right) \ge \sqrt{3 \beta}.
\end{equation}
where a minimal length is expected.

The most common approach to this deformed quantum mechanics is to keep the Hamiltonian unchanged, while to replace the position operator with a modified one. Written in momentum representation, the Schr\"odinger equation becomes a higher order differential equation. For example, to be consistent with the modified commutation relation \eqref{mcr}, the position operator should be
\begin{equation}
x = i\left( {1 + 3\beta {p^2}} \right)\frac{\partial }{{\partial p}},
\end{equation}
and the Schr\"odinger equation is modified correspondingly. With this method different systems with different potential have been studied. Kempf has given the results of the maximal localization states and harmonic oscillator in \cite{Kempf1995}. The energy levels of the hydrogen atom has been studied in \cite{Brau1999,Hossenfelder2003,Pedram2012}, Landau levels, Lamb shift, potential step and potential barrier in \cite{Ali2011}, and a single particle in a box in \cite{Ali2009,Nozari2006}. We note in passing that in fact different forms of modified commutation relation have been adopted in these works.

An elegant formulation in quantum physics is the path integral method. This method was applied to GUP deformed quantum mechanics and the free particle kernel was calculated in \cite{Das2012}. Inspired by their work, in this letter another useful approach--Schwinger's quantum action principle \cite{Schwinger1951}--was considered, and a comparison with \cite{Das2012} was presented.

In the next section we first consider a general theory until the applications are discussed. The modified commutation relations are given by
\begin{equation}
\left[ {{q_a}\left( t \right),{p_b}\left( t \right)} \right] = i{f_{ab}}
\end{equation}
where $f_{ab}$ is expected to be symmetric. Since we don't consider a minimal momentum, $f_{ab}$ is only a function of $p$.

\section{Schwinger's Quantum Action Principle}\label{qap}
As the name indicates, Schwinger's quantum action principle is an quantum mechanical analogy to the classical Hamilton's action principle. It is a variational approach to quantum mechanics and quantum field theory, closely related to Feynman's path integral formulation and equivalent to the equations of motion such as Schr\"odinger's equation and Heisenberg's equation. In its derivation, commutation relations are explicitly involved. We will see that the Hamilton's equations should be modified and respectively the quantum action principle. The derivations in \cite{Schwinger2001} were closely followed, and the difference induced by modified commutation relation was focused on. A quasi-position representation was adopted.

In the first place, with commutation relation modified two basic equations should be given by:
\begin{subequations}
\begin{align}
\frac{1}{i}\left[ {{q_a},F} \right] & = \frac{{\partial F}}{{\partial {p_b}}}{f_{ab}}\\
\frac{1}{i}\left[ {{p_a},F} \right] & =  - \frac{{\partial F}}{{\partial {q_b}}}{f_{ab}}
\end{align}
\end{subequations}

Since $f=f(p)$, the first equation always holds, while the second one is only approximate. For example, $\left[ {{p_a},{q_b}{q_b}} \right] = i\left( {{f_{ab}}{q_b} + {q_b}{f_{ab}}} \right) \ne 2i{q_b}{f_{ab}}$. For higher precision higher order expansions could be made, and the algebra would be quite complicated. To merely illustrate quantum action principle with GUP, we made a simplification here.

Using the Heisenberg's equations of motion we have
\begin{subequations} \label{Ham}
\begin{align} \label{Hama}
\frac{{d{p_a}\left( t \right)}}{{dt}} & = \frac{1}{i}\left[ {{p_a},H} \right]=-\frac{{\partial H}}{{\partial {q_b}}}{f_{ab}}\\
\frac{{d{q_b}\left( t \right)}}{{dt}} & = \frac{1}{i}\left[ {{q_b},H} \right]= \frac{{\partial H}}{{\partial {p_b}}}{f_{ab}}
\end{align}
\end{subequations}
which are no longer the elegant Hamilton's equations.

The Schr\"odinger equation leads to
\begin{equation} \label{sch}
\frac{\partial }{{\partial t}}\left\langle {q',t} \right| =  - i\left\langle {q',t} \right|H
\end{equation}

With the modified uncertainty relation the momentum operator in coordinate representation should be replaced by
\begin{equation} \label{partialq}
\langle q|{p_a} = {f_{ab}}\frac1{i} {\frac{\partial }{{\partial {q_b}}}\langle q|} \end{equation}

Substituting \eqref{sch} and \eqref{partialq} the change of the state when everything it depends on except the dynamics is infinitesimally varied is given by
\begin{multline}
{\delta _{{\rm{kin}}}}\langle q',t + dt|
= \frac{\partial }{{\partial {q_a}}}\langle q',t + dt|\delta {q_a}\left( {t + dt} \right) + \frac{\partial }{{\partial t}}\langle q',t + dt|\delta \left( {t + dt} \right) \\
= i\langle q',t + dt| \left( f_{ba}^{ - 1}(t+dt){p_b}\left( {t + dt} \right)\delta {q_a}\left( {t + dt} \right)
- H\left( {t + dt} \right)\delta \left( {t + dt} \right) \right)
\end{multline}

The change of the dual vector is obviously
\begin{equation}
{\delta _{kin}}\left| {q'',t} \right\rangle  =  - i\left( f_{ba}^{ - 1}\left(t\right){{p_b}\left( t \right) \delta {q_a}\left( t \right) - H\left( t \right)\delta t} \right)\left| {q'',t} \right\rangle
\end{equation}

Using the modified Hamilton's equations \eqref{Ham} yields
\begin{equation}
{p_a}\left( {t + dt} \right) = {p_a}\left( t \right) + dt\frac{d}{{dt}}{p_a}\left( t \right) = {p_a}\left( t \right) - {f_{ab}}\left(t\right)dt\frac{{\partial H\left(t\right)}}{{\partial {q_b}}}
\end{equation}

Now consider the kinematic variation of the transformation function
\begin{equation}
{\delta _{{\rm{kin}}}}\left\langle {q',t + dt{\kern 1pt} |{\kern 1pt} q'',t} \right\rangle  = i\left\langle q',t + dt\right|M\left| {q'',t} \right\rangle
\end{equation}
where
\begin{equation}
\begin{aligned}
M =&   f_{ab}^{ - 1}\left(t\right){p_b}\left( t \right)\left( {\delta {q_a}\left( {t + dt} \right) - \delta {q_a}\left( t \right)} \right)+df_{ab}^{-1}\left(t\right)p_b\left(t+dt\right)\delta q_a \left(t+dt\right)\\
&- dt\delta {q_a}\frac{{\partial H\left(t\right)}}{{\partial {q_a}}} - H\left(t\right)\delta dt - dt\frac{{\partial H\left(t\right)}}{{\partial t}}\delta t\\
 =&   f_{ab}^{ - 1}\left(t\right){p_b}\left( t \right)\left( {\delta {q_a}\left( {t + dt} \right) - \delta {q_a}\left( t \right)} \right) - \delta \left( {dtH\left(t\right)} \right) + f_{ab}^{ - 1}\delta {p_b}\frac{{d{q_a}}}{{dt}}dt\\
 &+df_{ab}^{-1}\left(t\right)p_b\left(t+dt\right)\delta q_a \left(t+dt\right)\\
 =&   f_{ab}^{ - 1}\left(t\right)\delta \left[ {{p_b}\left( t \right)d{q_a}\left( t \right)} \right] - \delta \left( {dtH\left( t \right)} \right) +df_{ab}^{-1}\left(t\right)p_b\left(t+dt\right)\delta q_a \left(t+dt\right)
 \end{aligned}
\end{equation}

Combining the dynamical change $\delta_{\rm{dyn}}\langle q',t+dt \, | \, q'',t \rangle =-i \langle q',t+dt \left| dt \delta H \right| q'',t \rangle$ leads to the total change
\begin{equation}
\delta \left\langle {{q',t + dt}}
 \mathrel{\left | {\vphantom {{q',t + dt} {q'',t}}}
 \right. \kern-\nulldelimiterspace}
 {{q'',t}} \right\rangle  = i\left\langle {q',t + dt} \right|\bar \delta W\left| {q'',t} \right\rangle
\end{equation}
where
\begin{equation}
\bar \delta W = f_{ab}^{ - 1}\left(t\right)\delta \left[ {{p_b}\left( t \right)d{q_a}\left( t \right)} \right] - \delta \left( {dtH\left( t \right)} \right)+df_{ab}^{-1}\left(t\right)p_b\left(t+dt\right)\delta q_a \left(t+dt\right)
\end{equation}
which is no longer a total change of operators. This inconsistency is from the approximation made above. The original counterpart is $\delta W= \delta \left[ dt \left( p_a \frac{dq_a}{dt} -H \right)  \right] = \delta \left( dt\, L\right)$.

With a bit calculation it can also be written as
\begin{equation}\label{deltaW}
\bar \delta W = f_{ab}^{ - 1}d\left( {{p_b}\delta {q_a}} \right) - d\left( {\delta tH} \right)+df_{ab}^{-1}\left(t\right)p_b\left(t+dt\right)\delta q_a \left(t+dt\right)
\end{equation}
which appeared as an alternative form and is sometimes more convenient for calculation.

\section{Application}
Let's see a simple example--the free particle kernel (propagator) studied in \cite{Das2012} where path integral approach was used. The one-dimensional modified commutation relation in their work is
\begin{equation}
\left[ x, p \right]= i \hbar \left( 1+ 3\beta p^2\right)
\end{equation}
or
\begin{equation}
f=   1+ 3\beta p^2
\end{equation}

We keep $\hbar$ in this section to make the dimensions clear. Consider two states $\langle x_2, t_2| \equiv \langle x_2|$ and $|x_1, t_1 \rangle \equiv |p_1\rangle$. The time inverval $t=t_2-t_1$ is infinitesimal.
In the equations below, $x_i$ could be the operator or its eigenvalue, and should be self-evident in the context.

The Hamiltonian is
\begin{equation}
H=\frac{p^2}{2m}
\end{equation}

The modified Hamilton's equations lead to
\begin{subequations}
\begin{align}
&{x_2} - {x_1} = \frac{{fpt}}{m}\\
&p_2-p_1=0
\end{align}
\end{subequations}

Substituting these equations in \eqref{deltaW} we get
\begin{equation} \label{change}
\begin{aligned}
\bar \delta W & = {f^{ - 1}}\left( {{p_2}\delta {x_2} - {p_1}\delta {x_1}} \right) - \delta t\frac{{{p^2}}}{{2m}} \\
& = {f^{ - 2}}\frac{{m\left( {{x_2} - {x_1}} \right)}}{t}\left( {\delta {x_2} - \delta {x_1}} \right) - \delta t\frac{{{f^{ - 2}}m{{\left( {{x_2} - {x_1}} \right)}^2}}}{{2{t^2}}}\\
 & \approx \left( {1 - 6\beta {p^2}} \right)\left[ {\frac{{m\left( {{x_2} - {x_1}} \right)}}{t}\left( {\delta {x_2} - \delta {x_1}} \right) - \delta t\frac{{m{{\left( {{x_2} - {x_1}} \right)}^2}}}{{2{t^2}}}} \right]
\end{aligned}
\end{equation}

The first equality sign is in fact accurate even though we use the result in Section \ref{qap}. This is true because in the derivation of the quantum action principle we have used the modified Hamilton's equation \eqref{Hama}, which is approximate if $f=f(p_a)$ does not commute with $q_b$, while fortunately in this simple case of free particle it vanishes.

In order to substitute $x_2$ and $x_1$ with their corresponding eigenvalues, the equation below is useful:
\begin{equation}
{\left( {{x_2} - {x_1}} \right)^2} = \mathcal{T}\left[ {{{\left( {{x_2} - {x_1}} \right)}^2}} \right] + \left[ {{x_2},{x_1}} \right] \approx x_2^2 - 2{x_2}{x_1} + x_1^2 - \frac{{it\hbar }}{m}
\end{equation}
where $\mathcal{T}$ is the time-ordering operator, and $f$ has been truncated to $1$, as we will do hereafter.

It is easy to see the the part $\left[ \ldots \right]$ in the last line of \eqref{change}:
\begin{multline}
\left\langle {{x_2}} \right|\left[ {\frac{{m\left( {{x_2} - {x_1}} \right)}}{t}\left( {\delta {x_2} - \delta {x_1}} \right) - \delta t\frac{{m{{\left( {{x_2} - {x_1}} \right)}^2}}}{{2{t^2}}}} \right]\left| {{x_1}} \right\rangle\\
= \left( {\delta \left( {\frac{{m{{\left( {{x_2} - {x_1}} \right)}^2}}}{{2t}}} \right) + \frac{{fi\hbar \delta t}}{{2t}}} \right)\left\langle {{{x_2}}}
 \mathrel{\left | {\vphantom {{{x_2}} {{x_1}}}}
 \right. \kern-\nulldelimiterspace}
 {{{x_1}}} \right\rangle
\end{multline}

Expand terms of $-6\beta p^2 \left[ \ldots \right]$, then we get
\begin{equation}
\begin{aligned}
& { - 6\beta {p^2}\left[ {\frac{{m\left( {{x_2} - {x_1}} \right)}}{t}\left( {\delta {x_2} - \delta {x_1}} \right) - \delta t\frac{{m{{\left( {{x_2} - {x_1}} \right)}^2}}}{{2{t^2}}}} \right]} \\
 &=   - 6\beta \frac{{{m^2}\left( {x_2^2 - 2{x_2}{x_1} + x_1^2 - \frac{{i\hbar t}}{m}} \right)}}{{{t^2}}} \left[ {\frac{{m\left( {{x_2} - {x_1}} \right)}}{t}\left( {\delta {x_2} - \delta {x_1}} \right) - m\delta t\frac{{x_2^2 - 2{x_2}{x_1} + x_1^2 - \frac{{i\hbar t}}{m}}}{{2{t^2}}}} \right]
 \end{aligned}
\end{equation}

Observing that $(x_2-x_1)^2$ always yields a term proportional to $\hbar$, after some straightforward calculations the only term which is proportional to $\hbar^2$ is
\begin{equation}
  - \frac{{3\beta {\hbar ^2}m\delta t}}{{{t^2}}},
\end{equation}
and the term proportional to $\hbar$ is
\begin{equation}
3i\beta {m^2}\hbar \delta \left( {\frac{{{{\left( {{x_2} - {x_1}} \right)}^2}}}{{{t^2}}}} \right).
\end{equation}

Combining these equations yields
\begin{equation}
\frac{{\delta \left\langle {{{x_2}}}
 \mathrel{\left | {\vphantom {{{x_2}} {{x_1}}}}
 \right. \kern-\nulldelimiterspace}
 {{{x_1}}} \right\rangle }}{{\left\langle {{{x_2}}}
 \mathrel{\left | {\vphantom {{{x_2}} {{x_1}}}}
 \right. \kern-\nulldelimiterspace}
 {{{x_1}}} \right\rangle }}
 =\delta \left( {\frac{{im{{\left( {{x_2} - {x_1}} \right)}^2}}}{{2t}} - \frac{1}{2}\ln t + \frac{{3i\beta \hbar m}}{t} - \frac{{3\beta {m^2}{{\left( {{x_2} - {x_1}} \right)}^2}}}{{{t^2}}} +  \ldots } \right)
\end{equation}
and thereby
\begin{equation}
 \left\langle {{{x_2}}}
  \mathrel{\left | {\vphantom {{{x_2}} {{x_1}}}}
 \right. \kern-\nulldelimiterspace}
 {{{x_1}}} \right\rangle
  = C\frac{1}{{\sqrt t }}\exp \left( {\frac{{im{{\left( {{x_2} - {x_1}} \right)}^2}}}{{2t}}} \right)
  \left( {1 + \frac{{3i\beta \hbar m}}{t} - \frac{{3\beta {m^2}{{\left( {{x_2} - {x_1}} \right)}^2}}}{{{t^2}}} +  \ldots } \right)
\end{equation}

The factor $C$ could be easily calculated to be $\sqrt{m/2\pi i \hbar}$.
The term $3i\beta \hbar m/t$ is identical to the result in \cite{Das2012},while the next term is only half of the corresponding term in their work. The deviation is from the truncation of $f$ in the last half part of our calculations.

\section{Conclusion}
Modifying the commutation relation for the introduction of a minimal length, we derived a modified Schwinger's quantum action principle and applied the result to a simple case--the free particle kernel. It was shown that the quantum action principle could be quite useful and convenient in some applications. Moreover, due to the importance of a propagator in path integral formulation and therefore the natural relation between the two approaches, an extension to path integral with GUP based on quantum action principle could be straightforward.

Due to the complexity introduced by the modified commutation relation, approximations in calculations are inevitable as in other formulations. However, as for the method itself, we are still looking  for a more concrete and consistent theory.

\end{document}